%% file: main.tex
\documentclass[amssymb,nobibnotes,superscriptaddress,longbibliography,notitlepage,twocolumn]{revtex4-1}

\usepackage{enumerate,amsmath}
\usepackage{graphicx}
\usepackage{color} 

\usepackage{braket}

\begin{document}

\title{Deep Variational Quantum Eigensolver: a divide-and-conquer method for 
solving a larger problem with smaller size quantum computers}%

\author{Keisuke Fujii}
\affiliation{Graduate School of Engineering Science, Osaka University, 1-3 Machikaneyama, Toyonaka, Osaka 560-8531, Japan.}
\affiliation{Center for Quantum Information and Quantum Biology, Institute for Open and Transdisciplinary Research Initiatives, Osaka University, Japan.}
\affiliation{RIKEN Center for Quantum Computing, RIKEN, Wako Saitama 351-0198, Japan}

\author{Kaoru Mizuta}
\affiliation{Department of Physics, Kyoto University, Kyoto 606-8502, Japan}

\author{Hiroshi Ueda}
\affiliation{Center for Quantum Information and Quantum Biology, Institute for Open and Transdisciplinary Research Initiatives, Osaka University, Japan.}
\affiliation{Computational Materials Science Research Team, RIKEN Center for Computational Science (R-CCS), Kobe, 650-0047, Japan}
\affiliation{JST, PRESTO, 4-1-8 Honcho, Kawaguchi, Saitama 332-0012, Japan}

\author{Kosuke Mitarai}%
\affiliation{Graduate School of Engineering Science, Osaka University, 1-3 Machikaneyama, Toyonaka, Osaka 560-8531, Japan.}
\affiliation{Center for Quantum Information and Quantum Biology, Institute for Open and Transdisciplinary Research Initiatives, Osaka University, Japan.}

\author{Wataru Mizukami}%
\affiliation{Center for Quantum Information and Quantum Biology, Institute for Open and Transdisciplinary Research Initiatives, Osaka University, Japan.}
\affiliation{JST, PRESTO, 4-1-8 Honcho, Kawaguchi, Saitama 332-0012, Japan}

\author{Yuya O. Nakagawa}
\affiliation{QunaSys Inc., Aqua Hakusan Building 9F, 1-13-7 Hakusan, Bunkyo, Tokyo 113-0001, Japan}

\begin{abstract}
We propose a divide-and-conquer method for the quantum-classical hybrid algorithm to solve larger problems with small-scale quantum computers. Specifically, we concatenate a
variational quantum eigensolver (VQE) with a reduction in the system dimension,
where the interactions between divided subsystems are taken as an effective Hamiltonian expanded by the reduced basis.
Then the effective Hamiltonian is further solved by VQE, which we call {\it deep VQE}. 
Deep VQE allows us to apply quantum-classical hybrid algorithms on small-scale quantum computers to large systems with strong intra-subsystem interactions and weak inter-subsystem interactions, or strongly correlated spin models on large regular lattices.
As proof-of-principle numerical demonstrations, we use the proposed method for quasi one-dimensional models, including one-dimensionally coupled 12-qubit Heisenberg anti-ferromagnetic models on Kagome lattices
as well as two-dimensional Heisenberg anti-ferromagnetic models on square lattices. The largest problem size of 64 qubits is solved by simulating 20-qubit quantum computers with a reasonably good accuracy $\sim$ a few $\%$.  The proposed scheme enables us to handle the problems of $>1000$ qubits by concatenating VQEs with a few tens of qubits.
While it is unclear how accurate ground state energy can be obtained for such a large system, our numerical results on a 64-qubit system suggest that deep VQE provides a good approximation (discrepancy within a few percent) and has a room for further improvement.
Therefore, deep VQE provides us a promising pathway to solve practically important problems
on noisy intermediate-scale quantum computers.
\end{abstract}

\maketitle
\section{Introduction}
Quantum computers are expected to solve certain problems,
such as prime factorization \cite{Shor1997}, quantum chemistry calculations \cite{AspuruGuzik2005, RevModPhys.92.015003}, 
and linear algebraic processes (matrix inversion) \cite{HHLalgorithm, Childs2017, Gilyn2019},
exponentially faster than classical computers.
By virtue of the extensive engineering effort paid for the realization of quantum computers,
we now have a quantum computer which is already intractable for classical computers to simulate,
namely quantum computing supremacy \cite{Arute2019}.
However, the size of current quantum computers is too small 
to implement fault-tolerant quantum computation,
where quantum information is protected by quantum error correction.
Such a transitional period is called noisy intermediate-scale quantum technology
(NISQ) era \cite{Preskill2018}.
Since the task of demonstrating quantum computing supremacy \cite{Arute2019, Bouland2018, Boixo2018} is not useful for practical applications,
our next milestone in the NISQ era is to demonstrate 
the advantage of using NISQ devices for those problems that expand our scientific frontier.

To this end, a significant amount of NISQ-oriented algorithms have emerged recently.
Among them, the variational quantum eigensolver (VQE) \cite{Peruzzo2014} has attracted much attention because of its notable feature that directly exploits quantum states generated on a quantum computer for practical problems such as quantum chemistry calculations.
While the objective of the original method was to find an approximate ground state of a quantum system, it has widely been extended since its first appearance.
Researchers have proposed various techniques, for example, to construct approximate excited states \cite{McClean2017, Nakanishi2019, Higgott2019, Parrish2019, Endo2019}, investigate nonequilibrium steady states in open quantum systems \cite{Yoshioka2019}, and calculate energy derivatives \cite{Mitarai2020, Parrish2019derivative, OBrien2019}. 

However, there are several serious problems in applications of real quantum devices:
noise is too high to perform deeper quantum computation,
and the number of qubits is too small to handle 
practically interesting problems.
Though we can resolve these by further experimental efforts
in the future, 
for the meantime  
we should develop algorithmic approaches to relax 
the hardware limitation. 
Regarding the noise issue,
error mitigation techniques \cite{Temme2017, Endo2018, Czarnik2020, Armands2020, Takagi2020, Cai2020},  has been investigated actively, 
and its experimental validity has already been demonstrated \cite{Kandala2019mitigation}.
To relax the hardware size or connectivity limitation, 
virtual quantum gates have been introduced 
to decompose a large quantum circuit into smaller ones 
with quasi-probability sampling \cite{Mitarai2019constructing, Mitarai2020overhead, peng2019simulating}.
There are several techniques for reducing the required number of qubits, for example, by exploiting symmetries of a target system \cite{bravyi2017tapering, setia2019reducing} or by so-called active space approximation \cite{McClean2017, Romero2018, Mizukami2019, Takeshita2020}.

For quantum chemical calculations that are considered a promising application of NISQ,  
various divide-and-conquer (DC) techniques have been developed. 
Say, a density-matrix DC approach or a fragmentation method is widely used to perform large scale molecular simulations~\cite{Yang1995DC,Gordon2012FMO}.
These methods are employed with density functional theory for weakly-correlated systems. 
Meanwhile, another set of methods, such as the cluster mean-field theory~\cite{Jimenez2015PRB,Hermes2019JCTC,Hermes2020JCTC}, multi-layer multiconfiguration time-dependent Hartree (ML-MCTDH)~\cite{Wang2003JCP,Meyer2003TCA}, active space decomposition techniques (ASD)~\cite{Parker2013JCP,Parker2014JCP,Nishio2019JCP}, quasi-complete-active-space (QCAS)~\cite{Nakano2000CPL}, the renormalization exciton model (REM)~\cite{AlHajj2005PRB,Zhang2012JCC}, and the $n$-body Tucker method~\cite{Mayhall2017JCTC,Abraham2020arXiv}, exist for quantum many-body systems with strong correlations in each subsystem and weak interactions between subsystems. This is by no means an exhaustive list, but the diversity and active development of DC methods reflect their importance in classical computing. NISQ has a severe limit on the number of available qubits considering the number of orbitals of a molecule. It is, therefore, highly desirable to develop a DC method designed in the framework of the quantum-classical hybrid algorithm. 

Here, we introduce a general framework for implementing a DC method on the quantum-classical hybrid algorithm,
which allows us to handle larger problems by diving them into small pieces
so that NISQ devices can solve practically important large problems.
It should be noted that while the use of DC techniques for the VQE has been explored in Ref. \cite{yamazaki2018practical,rubin2016hybrid}, where the authors proposed to combine existing DC techniques in the field of quantum chemistry~\cite{Yang1995DC,Gordon2012FMO,Knizia2012PRL}, this work provides a more general technique applicable to any quantum system consisting of subsystems with weak inter-subsystem interaction but strong intra-subsystem interaction.
To investigate the properties of such a system, 
we utilize multiple small-scale quantum computers that are
connected via classical computers.
We divide the system into small subsystems, 
each of which is solved, as the first step, by using VQE neglecting inter-subsystem interactions. 
The resultant approximated ground state is further used to 
generate a basis with reduced degrees of freedom to estimate an effective Hamiltonian
including the inter-subsystem interactions neglected in the first step.
We concatenate VQE to solve the effective Hamiltonian, which we call {\it deep VQE}.
In a sense, this scheme can be viewed as real-space renormalization using actual quantum devices.

We perform extensive numerical simulation on Heisenberg anti-ferromagnetic models with frustration as proof-of-principle demonstrations of deep VQE.
A quasi-one-dimensional system with 48 qubits in total can be tackled with 12-qubit quantum computers.
For a two-dimensional system,
we apply deep VQE for up to $8\times 8$ Heisenberg anti-ferromagnetic model on the square lattice using 16 or 20 qubit quantum computers.
As will be seen later, we successfully obtain a lower ground state energy than 
the energy calculated solely on the subsystems, which approaches the exact one.
Deep VQE will be a powerful approach to solving practically important problems on a quantum computer with a limited number of qubits.

\section{Deep VQE}
\subsection{Divide-and-conquer method for VQE}
\begin{figure}
\begin{center}
\includegraphics[width=0.9\linewidth]{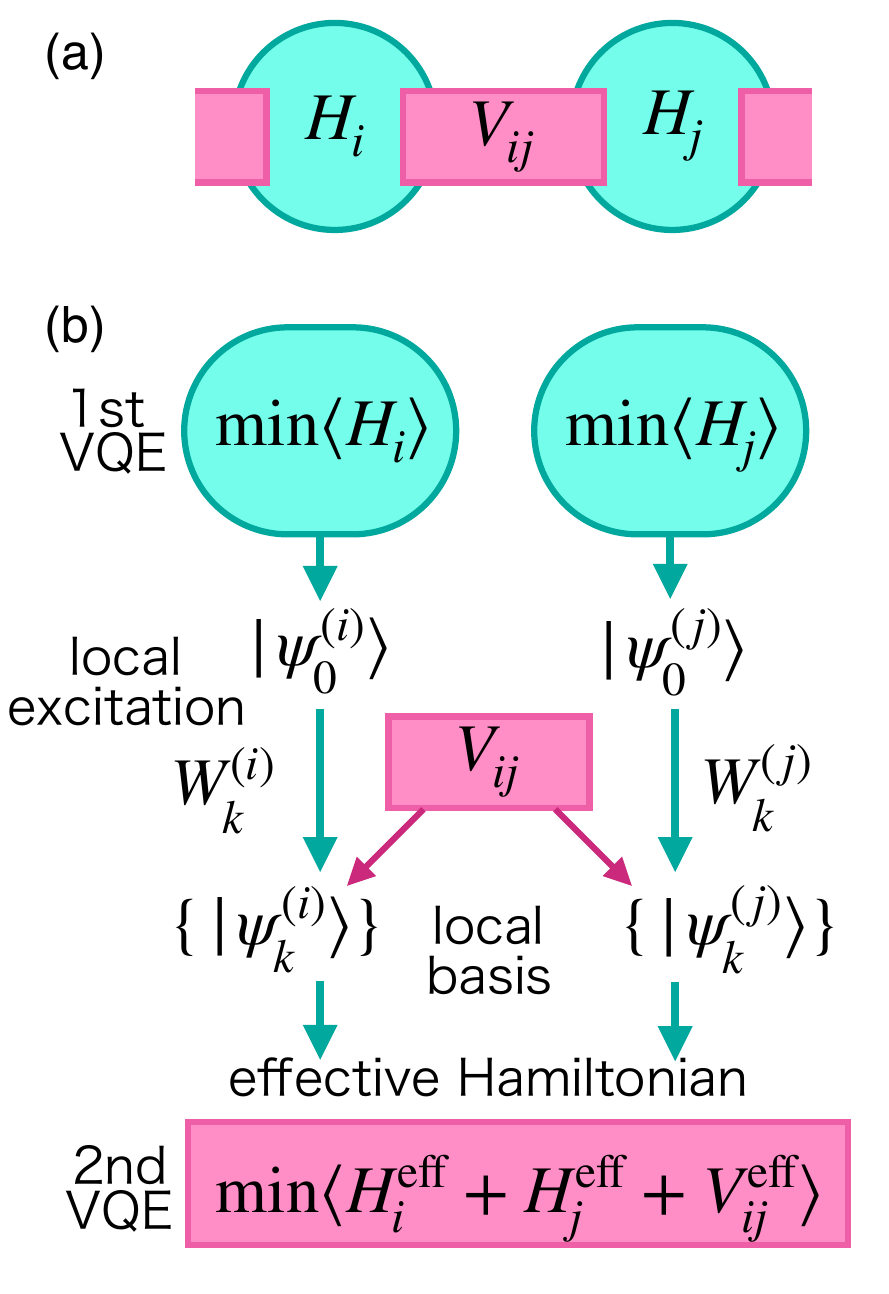}
\caption{ (a) The system consists of subsystems of Hamiltonian $H_i$, each of which interact with each other by inter-subsystem interaction $V_{ij}$.
(b) To solve the system depicted in (a), we first construct an approximate ground state $\ket{\psi^{(i)}_0}$ of each $H_i$ with VQE (the first VQE). Then we form a basis set by applying excitation operators on $\ket{\psi^{(i)}_0}$. Using the basis, we can construct an effective Hamiltonian which gives better approximation of the ground state.}
\label{fig1}
\end{center}
\end{figure}

Let us consider a Hamiltonian $H$,
which can be decomposed into a sum of subsystem Hamiltonian $H_i$ acting only on the $i$-th subsystem and interaction terms $V_{ij}$ acting on subsystems $i$ and $j$ [see Fig.~\ref{fig1} (a)]:
\begin{align}
H = \sum _{i} H_i + \sum _{ij} V_{ij}.
\end{align}
Suppose we have $N$ subsystems, each of which consists of $n$ qubits.
Let $M$ be the number of qubits required to describe the full Hamiltonian $H$, that is, $M=nN$.
The situation that we expect in this work is as follows:
each subsystem can be described by several tens/hundreds of qubits being subject to a strong intra-subsystem interaction, and 
these subsystems interact weakly with each other,
forming a larger system including thousands of qubits in total.
There are indeed many such systems at the molecular level, such as molecular aggregates,  molecular crystals, and dendrimers. 
Among those systems, this method would be suitable for describing singlet fission systems for organic light-emitting diodes (OLEDs)~\cite{Nagata2018AM} and solar cells~\cite{SingletFission2018CR}, or natural light-harvesting systems~\cite{NLHC2018RMP}.

Our idea here is to decompose such a problem 
into smaller problems.
As the first step, 
each subsystem Hamiltonian $H_i$
is solved by the conventional VQE with neglecting the inter-subsystem interactions,
which we call the first VQE below.
The qubits that are engaged in the inter-subsystem interactions 
are called a boundary of the subsystem.
The first VQE provides us a state close to the 
ground state of $H_i$, 
\begin{align}
| \psi^{(i)} _0 \rangle  =  U_i(\vec{\theta}^{(i),*}) | 0^n \rangle,
\end{align}
where $U_i ( \vec{\theta } )$ is a parameterized unitary circuit designed for $H_i$, and 
\begin{align}
\vec{\theta}^{(i),*} \equiv {\rm arg} \min _{\vec{\theta}^{(i)}} \langle 0^n | 
 U_i ({{\vec{\theta}}^{(i)}}){}^{\dag} H_i  U_i(\vec{\theta}^{(i)}) |0^n \rangle .
\end{align}
Hereafter we refer to $|\psi ^{(i)}_0 \rangle $ as a local ground state.

As the second step, we generate a $K$-dimensional local basis 
$\{ |\psi ^{(i)}_k \rangle \}_{k=1}^{K}$ from the local ground state by
\begin{align}
|\psi ^{(i)}_k \rangle \equiv W^{(i)}_k |\psi ^{(i)}_0 \rangle ,
\end{align}
where $\{ W^{(i)}_k\} $ is a set of operators 
on subsystem $i$, and $W^{(i)}_1$ is chosen to be an identity operator.
The operator $W^{(i)}_k$ $(k \neq 1)$ should be chosen to be a local excitation on 
a qubit at the boundary of the subsystem.
Suppose the inter-subsystem Hamiltonian 
is given by 
\begin{align}
V_{ij} = \sum _k v_{k} W^{(i)}_k W^{(j)}_k .
\label{eq:05}
\end{align}
Then the state is spanned by a product of the local basis
\begin{align}
\sum _k v_{k} \left( W^{(i)}_k|\psi ^{(i)}_0 \rangle \right)\left( W^{(j)}_k |\psi ^{(j)}_0 \rangle \right)
= V_{ij} |\psi ^{(i)}_0 \rangle|\psi ^{(j)}_0 \rangle ,
\end{align}
i.e., an entangled state with respect to the local bases,
contributes at least as a leading order correction of the perturbation theory
with a weak inter-subsystem interaction $V_{ij}$.
When the interaction term has a symmetry like a Heisenberg interaction, this entangled state recovers such a symmetry, even if it is broken in each subsystem.
A concrete choice of $\{W_k^{(i)}\}$, as an example, 
will be explained later. 

The overlap between basis states can be estimated 
as an expectation value of ${W_k^{(i)}}^{\dag} W_l^{(i)}$:
\begin{align}
\langle \psi ^{(i)}_k | \psi ^{(i)} _ l \rangle
= \langle 0 ^n |  U_i(\vec{\theta}^{(i),*})^{\dag} {W_k ^{(i)}}^{\dag} W_l^{(i)}  U_i(\vec{\theta}^{(i),*}) |0^n \rangle .
\end{align}
Since  $W_k^{(i)}$ is a local excitation,
${W_k ^{(i)}}^{\dag} W_l^{(i)}$ can be decomposed into a finite number of Hermitian operators.
This allows us to calculate the overlap
without any indirect measurement.
Using the above inner product, 
we can also define 
the orthonormal basis 
$\{ |\tilde \psi ^{(i)}_k \rangle \}$
using Gram-Schmidt process:
\begin{align}
|\tilde \psi ^{(i)}_k \rangle =
\frac{1}{C}\left(| \psi ^{(i)}_k \rangle  -  \sum _{l<k} \langle \tilde \psi ^{(i)}_l | \psi ^{(i)}_k \rangle |\tilde \psi ^{(i)}_l \rangle  \right),
\end{align}
where $C$ is the normalization factor which can be calculated from $\{ \langle \psi ^{(i)}_k | \psi ^{(i)} _ l \rangle \}$.
In this way, 
two bases are related by 
a $K \times K$ matrix $P^{(i)}$,
\begin{align}
|\tilde \psi _k ^{(i)} \rangle 
= \sum _{k'=1}^{K} P^{(i)}_{kk'} |\psi _k ^{(i)} \rangle ,
\end{align}
where the matrix element $P^{(i)}_{kk'}$ can be obtained from 
$\{ \langle \psi ^{(i)}_k | \psi ^{(i)} _ l \rangle \}$.
Hereafter, we simply call this orthogonal basis a local basis.

At the third step, 
the effective Hamiltonian is constructed using the local basis.
For the subsystem Hamiltonian,
the matrix representation of the effective Hamiltonian with respect to the local basis 
is defined as follows:
\begin{align}
(H_i ^{\rm eff})_{kl} = \langle \tilde \psi ^{(i)}_k | H_i | \tilde \psi ^{(i)}_l \rangle .
\end{align}
Note that since the state that we can easily generate
is $|\psi _k^{(i)}\rangle$,
the effective Hamiltonian is calculated from 
\begin{align}
(\bar H_i ^{\rm eff})_{kl} = \langle \psi ^{(i)}_k | H_i | \psi ^{(i)}_l \rangle
\end{align}
in actual calculations.
Similarly to the previous case,
$(\bar H_i ^{\rm eff})_{kl}$ can be estimated 
with direct measurements
by decomposing 
\begin{align}
     {W_k ^{(i)}}^{\dag} H_i W_l^{(i)} 
\end{align}
into a linear combination of Hermitian operators,
whose number is proportional to the number of terms in $H_i$.
$(\bar H_i ^{\rm eff})_{kl}$ and $(H_i ^{\rm eff})_{kl}$
are related by 
\begin{align}
(H_i ^{\rm eff})_{kl}  = 
\sum _{k'l'} {P^{(i)}}^{*}_{kk'} (\bar H_i ^{\rm eff})_{k'l'} P^{(i)}_{ll'}
=\left( {P^{(i)}}^{*} \bar H_i ^{\rm eff} {P^{(i)}}^{\rm T} \right)_{kl},
\end{align}
where $*$ and ${\rm T}$ indicate complex conjugate and transpose, respectively.

In addition, we take
the inter-subsystem interactions,
which are neglected in the first step.
Their matrix representations are defined by using the 
associated local bases:
\begin{align}
(V_{ij}^{\rm eff})_{kk'll'}
= \langle \tilde \psi^{(i)}_k | \langle \tilde \psi ^{(j)}_{k'} | V_{ij} | \tilde \psi ^{(i)}_l \rangle  | \tilde \psi ^{(j)}_{l'} \rangle .
\end{align}
Recall that the interaction term $V_{ij}$ defined in Eq.~(\ref{eq:05})
is written as a sum of tensor product operators.
Then it can be estimated 
by using an $n$-qubit quantum computer from 
\begin{align}
(V_{ij}^{\rm eff})_{kk'll'}
=  \sum _{\nu} v_\nu  \langle \tilde \psi^{(i)}_k | W^{(i)}_\nu | \tilde \psi ^{(i)}_l \rangle
\langle \tilde \psi ^{(j)}_{k'} | W_\nu ^{(j)} | \tilde \psi ^{(j)}_{l'} \rangle.
\end{align}
Note that it is enough to calculate the matrix elements independently 
on each subsystem.
Similarly to the previous case,
the effective inter-subsystem Hamiltonian 
is calculated from expectation values obtained by $\{ |\psi _k ^{(i)}\rangle\}$
applying the linear transformation $P^{(i)}$.
In this way, we now have 
an effective Hamiltonian $H^{\rm eff}$ of $H$,
\begin{align}
H^{\rm eff} = \sum _{i} H^{\rm eff}_i + \sum _{ij} V^{\rm eff}_{ij},
\end{align}
which acts on the $K^N$-dimensional system.
For a fixed accuracy, 
this takes $O({\rm poly}(M) K^4 N)$ runs of quantum computers of $n$ qubits,
where ${\rm poly}(M)$ is responsible for 
counting the number of terms in $H$.
The energy expectation value with respect to a product state of the local basis, $\bigotimes_{i=1}^N\ket{\tilde \psi^{(i)}_0}$, can be written as, 
\begin{align}
H^{\rm eff}_{00} := \sum _{i} (H^{\rm eff}_i)_{00} + \sum _{ij} (V^{\rm eff}_{ij})_{0000},
\end{align} 
which is the starting point of improving 
the ground state energy in the proposed scheme.

As the fourth step, which is crucial in the proposed scheme, 
we use VQE again to find the ground state of the 
effective Hamiltonian.
Suppose we have an $m$-qubit system, 
where $m$ is chosen to be $m= N \lceil \log_2 (K) \rceil$.
The number of qubits is reduced from $M$ to $m$.
A parameterized quantum circuit $V(\vec{\phi})$ to generate an approximate ground state of $H^{\mathrm{eff}}$
is constructed appropriately so that 
$V(\vec{\phi})$ acts on $K^N$-dimensional subspace of the $2^m$-dimensional Hilbert space.
Then the expectation value of the effective Hamiltonian 
can be expressed as:
\begin{align}
\langle 0^m|V(\vec{\phi}) ^{\dag} H^{\rm eff} V(\vec{\phi})|0^m \rangle,
\end{align}
which serves as the cost function of the second VQE.
Note that if the ground state energy is set to be negative, 
a parameterized quantum circuit acting fully on the $m$-qubit system 
finds the ground state in the $K^N$-dimensional subspace appropriately,
simply by minimizing the energy expectation value.

The effective Hamiltonian $H^{\rm eff}$, which is described by $K\times K$ 
and $K^2 \times K^2$  dense matrices $(H_i ^{\rm eff})_{kl}$ and $(V_{ij}^{\rm eff})_{kk' ll'}$ respectively, 
can be written as a linear combination of at most $O[{\rm poly}(M)K^4]$ 
$m$-qubit Pauli operators,
and hence can be estimated by $O[{\rm poly}(M)K^4]$ runs of quantum computers of $m$ qubits.
By minimizing the cost function, 
we obtain a better approximation of the 
ground state and its energy.
One might think that the accuracy of estimating the matrix elements of  $H^{\rm eff}$  would have a significant impact on the accuracy of the final energy, but this is not the case.
According to a matrix perturbation theory, 
even if each element in a matrix has an additive error $\epsilon$,
the corresponding accuracy of the energy eigenvalue is bounded by ${\rm poly} (M) K^2 \epsilon$.
More precisely, 
suppose the estimated Hamiltonian $\tilde H^{\rm eff}$ is given by 
\begin{align}
    \tilde H^{\rm eff} = H^{\rm eff} + H^{\rm error},
\end{align}
where the absolute value of each element of $H^{\rm error}$ 
is bounded by $\epsilon$.
Then the corresponding energy eigenvalues $\tilde E$ and $E$ satisfy
\begin{align}
    |\tilde E- E | \leq \| H^{\rm error} \|_{\infty}.
\end{align}
Since each term of $H^{\rm error}$ (and also $\tilde H^{\rm eff}$) has a tensor product structure so as to act on at most $K^2$-dimensional subsystems. Therefore we have
\begin{align}
    |\tilde E- E | \leq \| H^{\rm error} \|_{\infty} \leq {\rm poly}(M) K^2 \epsilon .
\end{align}
Therefore, the accuracy can be guaranteed efficiently if $\epsilon$ is sufficiently small. 
Let $\bar \epsilon$ be a target accuracy of the energy.
Then,  $\epsilon$ should be 
\begin{align}
    \epsilon = \frac{\bar \epsilon}{{\rm poly}(M)K^2 },
\end{align}
and hence the number of measurements should be scaled $1/\epsilon ^2={\rm poly}(M)K^4 / \bar{\epsilon}^2$.
\begin{figure}[t]
\begin{center}
\includegraphics[width=0.95\linewidth]{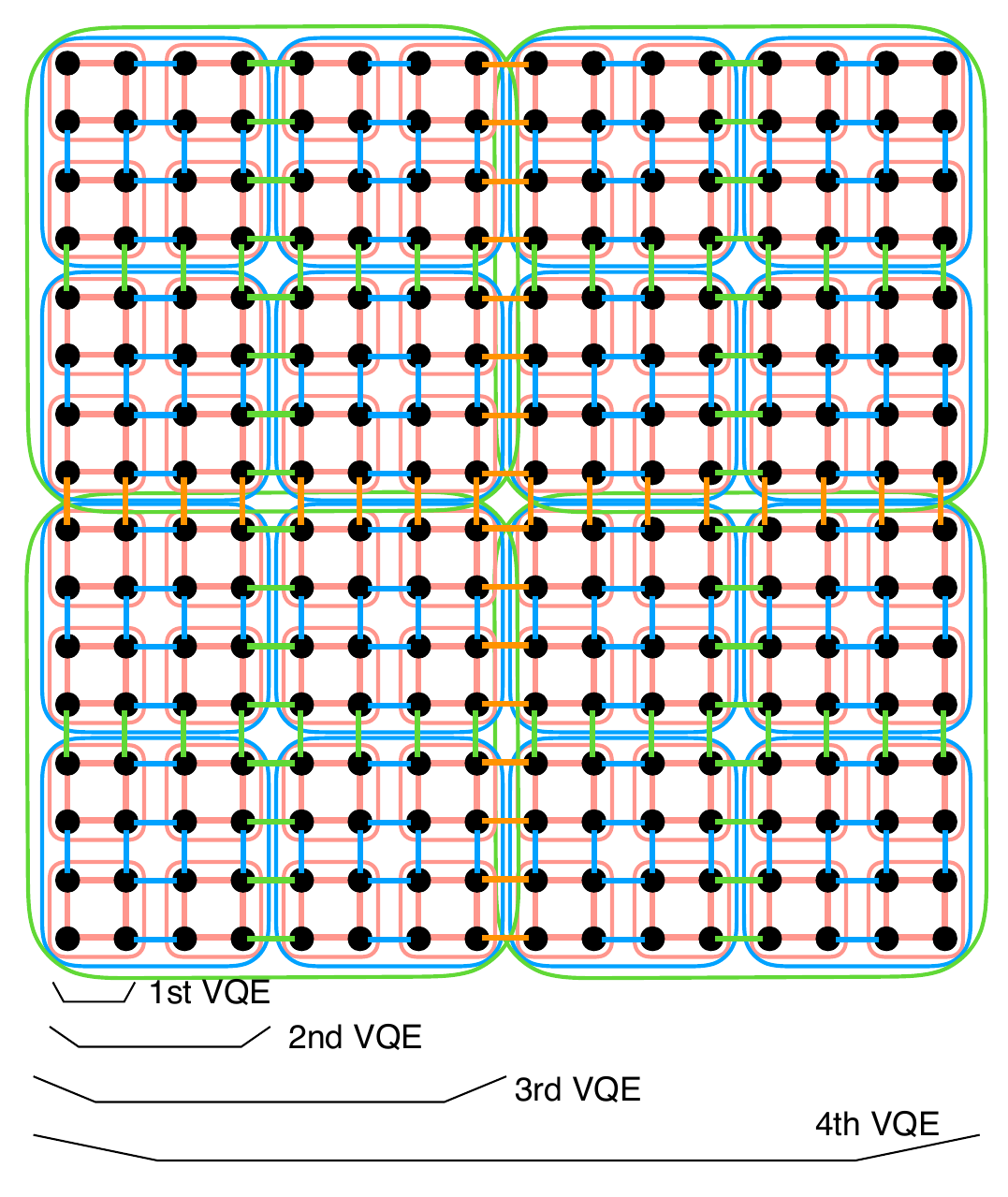}
\caption{Concatenation of VQEs. Red, blue, green squares correspond to the 1st, 2nd, and 3rd VQEs, respectively. 
The red, blue, green, and orange edges indicate the inter-subsystem interactions taken at the 1st, 2nd, 3rd, and 4th VQEs.
At each level, the local basis is generated by the local excitations on each qubit at the boundary, that is, the qubits engaged in the inter-subsystem interactions. At each level, the effective Hamiltonian is constructed from suitably chosen local basis. }
\label{fig1b}
\end{center}
\end{figure}

Note that the proposed scheme shares an idea with Ref.~\cite{subspaceexpansion} to estimate 
the effective Hamiltonian from the approximated ground state 
with local excitations. 
However, here we crucially put the step 
forward; the effective Hamiltonian is constructed including 
the interactions that are neglected when dividing the system into subsystems,
and the effective Hamiltonian with reduced degrees of freedom 
is further solved by VQE at the second stage.

We also comment that
instead of generating the local basis by a local excitation $W_k$,
we can use subspace-search VQE to find an orthogonal basis of a low energy subspace~\cite{Nakanishi2019}.
However, we find that this low energy expansion results in 
worse energy than the above construction when the same number of dimensions of the local basis is employed. This might be attributed to the boundary error 
in the real-space renormalization as mentioned in Ref.~\cite{DMRGreview}.
Hopefully, the local excitations at the boundary 
can handle this issue, at least in a perturbative way as mentioned previously.

\subsection{Multiple concatenations of VQE}
In the above explanation, 
we concatenated VQE only twice.
However, the procedure can be executed recursively 
to make a hierarchical structure to divide a larger problem into smaller pieces, where
the correlations are taken like a real space renormalization
as shown in Fig.~\ref{fig1b}.
Suppose a two-dimensional system is divided into 
multiple subsystems consisting of $l^{(1)} \times l^{(1)}$ qubits. 
After the first VQE, the local basis can be generated by the local excitations at the boundary.
The dimensions of the local basis scale like $O(l^{(1)})$. Even if we take local excitations 
for all qubits including the bulk, the dimensions of the local basis are only ${\rm poly}(l^{(1)})$. 
This means that, in the second level, the subsystem can be 
handled by $O(\log_2 (l^{(1)}))$ qubits. 
By using the local basis, the effective Hamiltonian of the first level $H_{\rm eff}^{(1)}$ is constructed
including the inter-subsystem interactions that are neglected in the first stage.
In addition, we obtain the effective expression $\{ W_{k,{\rm eff}}^{(1)}\}$ of the local excitations at the boundary $\{ W_k \}$, to generate the local basis 
in the next level.

In the second level, 
we consider $l^{(2)} \times l^{(2)}$ lattice, each site of which is the system solved in the first VQE.
The second VQE requires only $O[(l^{(2)})^2 \log(l^{(1)})]$ qubits. 
The local basis is generated 
by applying the local excitations $\{W_{k, {\rm eff}}^{(1)}\}$
at the boundary of each subsystem in the second level.
The dimensions of the local basis, i.e., the number of local excitations, are 
proportional to the 
length $O(l^{(1)} l^{(2)})$ of the boundary at the lowest level.
By using the second-level local basis, the effective Hamiltonians 
and local excitations are constructed similarly.

By recursively repeating this procedure,
at the $(k-1)$th level, 
the state obtained by the $(k-1)$th VQE is used to 
generate a local basis. 
The length of the boundary at the lowest level is 
\begin{align}
l_{k-1} \equiv \prod_{j=1}^{k-1} l^{(j)},
\end{align}
and hence the dimensions of the local basis are ${\rm poly}(l_{k-1})$.
By using the local basis, the effective Hamiltonian and local excitations
at the $k$th level is obtained.
At the $k$th-level concatenation, 
$l^{(k)} \times l^{(k)}$ lattice, where each site corresponds 
to the system spanned by the local basis in the $(k-1)$th level, 
is solved by $k$-th VQE.
The number of qubits required in the $k$th level is 
$O[(l^{(k)})^2 \log(l_{k-1})]$. 
Since the number of qubits handled at the lowest level increases exponentially 
in the number of concatenation $k$,
it is enough to choose $k$ as a logarithmic function of the problem size, i.e., the total number of physical qubits $M$.
Then the total number of runs of quantum computers is only a polynomial in the problem size $M$.
The number of qubits required is only logarithmic, $O[(l_{\rm max})^2 \log(M)]$, in the problem size $M$,
where $l_{\rm max} = \max_k l^{(k)}$ is chosen to be a constant.
In principle, this procedure can accommodate entanglement entropy scaling like $O[\log(|\partial D|)]$,
where $|\partial D|$ is the length of the boundary of a region $D$.

In Fig.~\ref{fig1b},
we show the case with $l^{(k)}=2$, where concatenation is performed up to $k=4$ with a periodic boundary condition.
Suppose three types of local excitations, for example, corresponding to the Pauli operators, are introduced on each qubit at the boundary.
The dimensions of the local bases are $K=13$, 37 and 85 at the 2nd, 3rd and 4th level, respectively,
which means that we should use
$4$, $6$, and $7$ qubits to represent each subsystem.
In this case, 4, 16, 24, and 28 qubits in total are employed in each of 1st, 2nd, 3rd, and 4th VQEs, respectively.
The total number of physical qubits is 256.
If we add one more concatenation,
a square lattice of length 32, i.e., 1024-qubit systems can be handled with 32-qubit quantum computers.

\begin{figure*}[t]
\begin{center}
\includegraphics[width=0.95\linewidth]{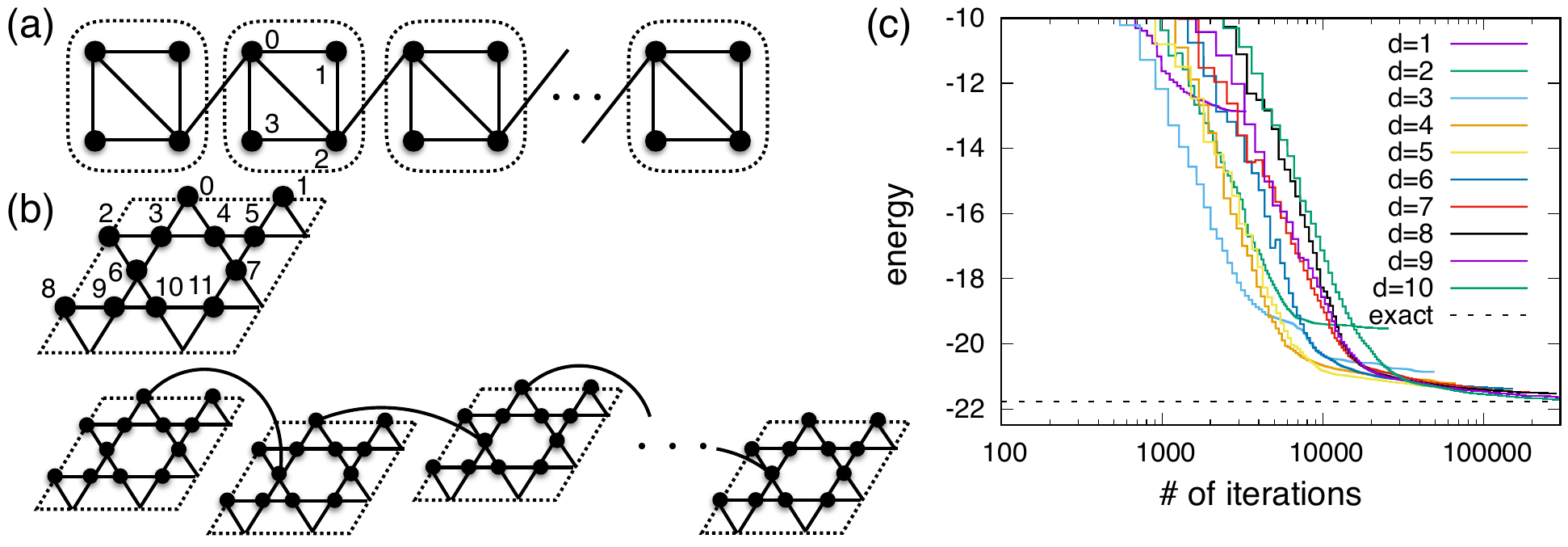}
\caption{ (a) The system used for the numerical demonstration. It consists of 4-qubit subsystems of Hamiltonian $H_i$, each of which interacts with each other by inter-subsystem interaction $V_{ij}$. (b) A unit cell of Heisenberg anti-ferromagnetic model on a 12-qubit Kagome lattice. The 12-qubit systems interact with each other in a nearest-neighbor way. (c) The energy obtained by the first VQE for the 12-qubit Heisenberg anti-ferromagnetic model. $d$ indicates the depth, i.e., the number of cycles, of the parameterized quantum circuits. See tab.~\ref{Tab_for_fig3} for the converged energies with increasing the ansatz depth.\label{fig2}}
\end{center}
\end{figure*}

\begin{table*}[]
\begin{center}
\caption{
\label{Tab_for_fig3} Convergence of the energy for 12-qubit Kagome lattice with increasing ansatz depth.}
\begin{tabular}{c|cccccccccc}
\hline \hline
depth $d$ & 1 & 2 & 3 & 4 & 5 & 6 & 7 & 8 & 9 & 10
\\
\hline
energy  & $-12.86$ & $-19.53$ & $-20.96$ & $-21.21$ & $-21.43$ & $-21.38$ & $-21.48$ & $-21.53$ & $-21.65$ & $-21.72$
\\
\hline \hline
\end{tabular}
\end{center}
\end{table*}%

\section{Numerical simulation}
\subsection{Quasi one-dimensional systems}
To make the proposed scheme more concrete, 
we will demonstrate a series of numerical simulations.
Numerical simulations are done by using Qulacs, 
an open source fast quantum computer simulator on classical computers~\cite{Qulacs}.
First, we consider the case where 
each local subsystem is governed by a 4-qubit 
Heisenberg anti-ferromagnetic model:
\begin{align}
H_i = \sum _{(\mu,\nu) \in E} (X^{(i)}_\mu X^{(i)}_\nu + Y^{(i)}_\mu Y^{(i)}_\nu + Z^{(i)}_\mu Z^{(i)}_\nu),
\end{align}
where the Pauli operator $A^{(i)}_\mu$ with $A \in \{X,Y,Z\}$ indicates 
the Pauli operator acting on the $\mu$th qubit in subsystem $i$,
and $E=\{ (0,1), (1,2), (2,3), (3,0), (0,2)\}$ is a set of edges.
The whole system consists of $N$ such subsystems that are coupled in a one-dimensional way via
a Heisenberg anti-ferromagnetic interaction as shown in Fig.~\ref{fig2} (a):
\begin{align}
V_{ij} = X^{(i)}_0 X^{(j)}_2 + Y^{(i)}_0 Y^{(j)}_2 + Z^{(i)}_0 Z^{(j)}_2,
\end{align}
where the 0th qubit in the $i$th subsystem and the 2nd qubit in the $j$th subsystem are engaged in the interaction.

The parameterized quantum circuit is constructed as follows.
For each cycle, 
we apply an arbitrary single-qubit gate on each qubit
followed by a two-qubit gate generated by 
the Heisenberg interaction
\begin{align*}
X_{\mu}X_{\nu}+Y_{\mu}Y_{\nu}+Z_{\mu}Z_{\nu},
\end{align*}
on each edge in $E$ in a certain order.
The cycle is repeated several times.
The rotational angles with respect to the Pauli operators and the Heisenberg interactions
are treated as the parameters of single-qubit and two-qubit gates, respectively.
The parameters are optimized by using BFGS (Broyden–Fletcher–Goldfarb–Shanno) by using numerical differentiation.
In an actual experiment, the gradient of the parameters should be obtained by using 
the parameter shift rule~\cite{mitarai2018}. 
The exact ground state energy of $H_i$ is $-7.0$.
The VQE with two cycles provides us an exact ground state $|\psi _0^{(i)}\rangle$ with the energy of $-7.0$ with the fidelity 1.0.
This allows us to separate the performance analysis of 
the proposed scheme below from 
the imperfection of VQE at the first stage.
Then, in addition to $|\psi _0^{(i)}\rangle$, we generate a local basis by
the Pauli operators engaged in the inter-subsystem interactions:
\begin{align}
\{| \psi _0 ^{(i)} \rangle, A^{(i)}_0 | \psi _0 ^{(i)} \rangle , \;\;\;  A^{(i)}_2 | \psi _0 ^{(i)} \rangle \} . 
\end{align}
In this case, the dimension of the local basis is $K=7$,
which can be treated with three qubits.
While the dimensional reduction is not so large in this case, 
we regard this task as a validation of the proposed scheme.
We calculate the effective Hamiltonian $H^{\rm eff}$
and solve it again with the VQE.
In the second VQE, 
we use a parameterized quantum circuit which is constructed 
a single-subsystem unitary gate generated
by the effective subsystem Hamiltonian $H_i^{\rm eff}$ 
and two-subsystem unitary gate generated by $V_{ij}^{\rm eff}$ with their rotational angles are taken as the parameters:
\begin{align}
U(\vec\phi)  &\equiv \prod _{l} W_l (\vec\phi ^{(l)}),
\\
W_l  (\vec\phi ^{(l)}) & \equiv \prod _{i} e^{- i \phi ^{(l)}_i H_i^{\rm eff}} \prod _{jk} e^{- i \phi^{(l)} _{jk} V_{jk}^{\rm eff}}.
\end{align}
Note that these gates, $e^{- i \phi ^{(l)}_i H_i^{\rm eff}}$
and $e^{- i \phi^{(l)} _{jk} V_{jk}^{\rm eff}}$, are $\lceil \log(K) \rceil$-qubits and 
$\lceil \log_2(K^2) \rceil$-qubits gate, respectively.
Such unitary gates can be compiled from elementary single-qubit gates 
and two-qubit gates by Solovay-Kitaev algorithm or variational quantum gate optimization~\cite{VQGO}.
These work efficiently, since $K$ is chosen to be at most polynomially large in the problem size $n$.

The results are summarized in Tab. ~\ref{Tab1},
where the energy expectation value calculated from 
the product state of the local ground state $|\psi _0 \rangle^{\otimes N}$, 
the exact ground energy for $H^{\rm eff}$, and the exact ground state 
energy of $H$ estimated by 
the Lanczos method 
on Qulacs or QS$^3$
are also shown as a comparison.
\begin{table}[t]
\begin{center}
\caption{
\label{Tab1}Numerical results for $4 \times N$ Heisenberg anti-ferromagnetic systems shown in Fig.~\ref{fig2} (a). ``Deep VQE" indicates the results obtained by the proposed scheme. ``Local" indicates the energy calculated from a product state of the local ground state $|\psi _0 \rangle $, i.e., $H^{\rm eff}_{00}$. ``Effective" means the exact ground state energy of $H^{\rm eff}$. ``Exact" is a ground state energy calculated by Lanczos method on a simulator; the exact energy for $4 \times 8$ system is computed by use of a quantum spin solver QS$^3$\cite{2021arXiv210700872U}}.
\begin{tabular}{c|cccc}
\hline \hline
system  & Deep VQE & Local & Effective & Exact
\\
\hline
$4\times 2$ & $-14.46$    & $-14.00$ & $-14.46$ & $-14.46$ 
\\
$4 \times 3$ & $-21.89$ & $-21.00$ & $-21.89$  &$ -21.92$
\\
$4\times 4$ & $-29.31$    & $-28.00$ & $-29.32$ & $-29.39$
\\
$4\times 5$ & $ -36.70$   & $-35.00 $& $-36.75 $  &    $-36.85$
\\
$4\times 6$ & $-44.13$    & $-42.00$ & NA & $ -44.31$
\\
$4\times 8$ & $-59.02$    & $-56.00$ &NA & $-59.23$
\\
\hline \hline
\end{tabular}
\end{center}
\end{table}%
In the case of $N=2$, 
the proposed method provides the almost exact ground state.
In the case of 
$3 \leq N \leq 8$, the obtained energies are 
$0.1$\%-$0.4$\%
higher than the exact energy obtained by 
the Lanczos method.
This is attributed to the fact that the local basis employed to expand $H^{\rm eff}$ is not enough to achieve an exact ground state energy,
since VQE at the second stage successfully provides
the ground state of $H^{\rm eff}$.
In all cases, 
we can see that the proposed scheme provides a better approximation of the ground state energy 
smaller than those obtained by local ground states, i.e., $H^{\rm eff}_{00}$.
This implies that an entangled state of local basis states 
is generated to reduce the total energy.

In the above example, the effect of dimensional reduction is 
small.
Next, we consider a tougher example,
where each local subsystem is a Heisenberg anti-ferromagnetic model with a 12-qubit Kagome lattice, as shown in Fig~\ref{fig2}~(b).
The parameterized quantum circuit is constructed in the same way as the previous case.
The first VQE with depth 10 for the 12-qubit subsystem
results in a good approximation $-21.72$ 
of the exact ground state energy $-21.78$,
which corresponds to fidelity 0.977 as shown in Fig.~\ref{fig2} (c) and Tab.~\ref{Tab_for_fig3}.
The inter-subsystem interactions are introduced so that 
the 12-qubit subsystems interact
in a one-dimensional way. Specifically, 
0th and 6th qubits, each of which belongs to 
neighboring subsystems, interact with the Heisenberg 
anti-ferromagnetic interaction.
The local basis is generated in the same way as the previous example.
This means that we approximate the 12-qubit system as a 7-dimensional system, i.e., three qubits,
and hence the dimensional reduction enabled by 
the proposed method is apparent.
The ansatz for the second VQE is again constructed in the same way as the previous case.

In the case of $N=2$ and $4$, i.e., two and four subsystems, respectively, 
the proposed method results in energy $-43.8$ and $-87.9$,
both of which achieve the exact ground state energy of 
the effective Hamiltonians. 
In the case of $N=2$, 
the Lanczos method provides 
-44.055 and the deep VQE works well while reducing the total number of qubits.
While in the case of $N=4$ we cannot compare the result with the exact energy obtained 
by 
the Lanczos method
, at least we can say that 
the energy obtained is smaller than the energy 
expectation value $H^{\rm eff}_{00}=-43.4$ and $H^{\rm eff}_{00}=-86.9$ estimated by a product state of 
the local ground state $|\psi _0 \rangle$. 
In this case, the problem of the 48-qubit system
is solved using VQEs with 12 qubits,
which are feasible within the current technology.
The accuracy would be improved by appending more states to the local basis. Even if the dimensions of the local basis are doubled,
it only results in adding one more qubit to each site 
in the second VQE.

\subsection{Two-dimensional systems}

\begin{figure*}[t]
\begin{center}
\includegraphics[width=0.95\linewidth]{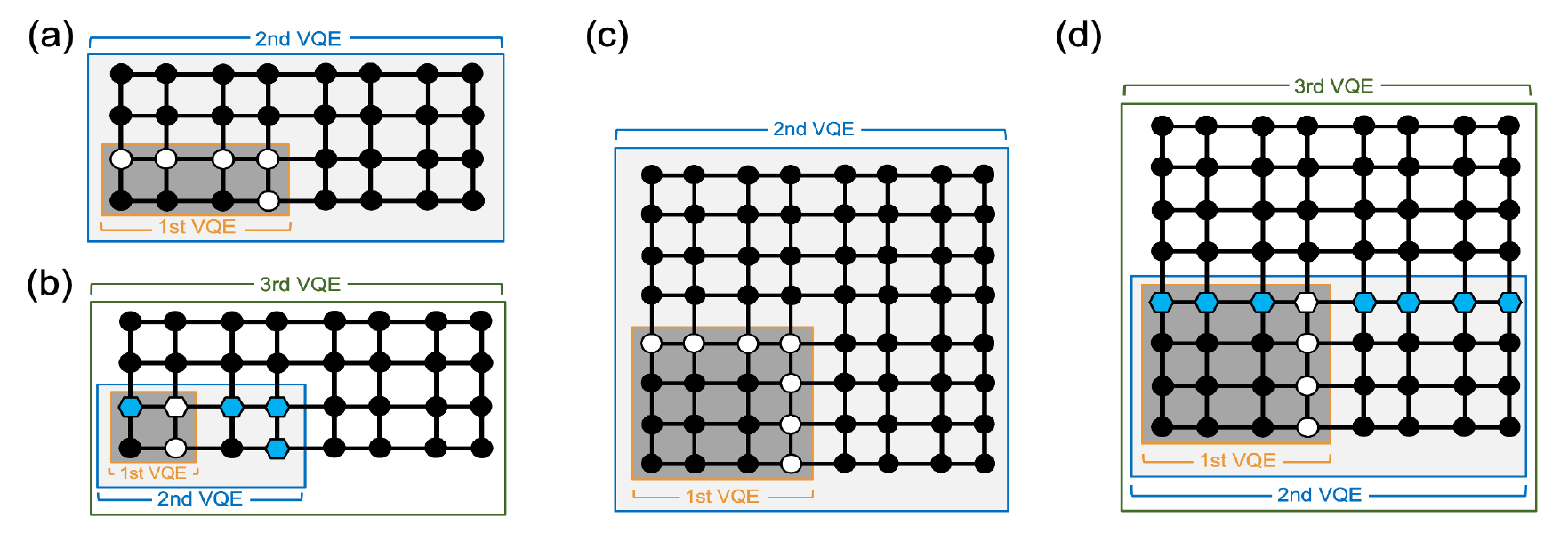}
\caption{ The protocol of concatenating of VQEs employed in the numerical demonstration of 2d systems. Qubits marked by white symbols are located at the boundary of subsystems for the 1st VQE. Hexagons (both blue and white) indicate the boundary sites of subsystems for the 2nd VQE, which are used for constructing the local basis following Eq. (\ref{eq:local_basis_b}). (a) $32$-sites with the concatenation up to the 2nd VQE. (b) $32$-sites with the concatenation up to the 3rd VQE. (c) $64$-sites with the concatenation up to the 2nd VQE. (d) $64$-sites with the concatenation up to the 3rd VQE.}
\label{fig3}
\end{center}
\end{figure*}

We also demonstrate the performance of Deep VQE and the further concatenated VQE for two-dimensional systems, which give rise to larger reduction of qubits. We consider a Heisenberg anti-ferromagnetic model on a two-dimensional (2d) $(L_x \times L_y)$-site square lattice, whose Hamiltonian is
\begin{align}\label{eq_hamiltonian2d}
    H = \sum_{ \{\mu, \nu \} \in E} (X_\mu X_\nu + Y_\mu Y_\nu + Z_\mu Z_\nu).
\end{align}
The set of edges $E$ is composed of pairs of neighboring sites under the open boundary condition. Assuming that both $L_x$ and $L_y$ are finite and even, this model has a unique ground state respecting the SU(2) symmetry.

We examine our protocol up to the 2nd VQE on a $16$-qubit system with $L_x=4,L_y=4$. We choose each subsystem for the 1st VQE by a $(2 \times 2)$-site lattice (See Fig. \ref{fig1b}). The 1st VQE is performed by a hardware-efficient ansatz with depth $10$, reproducing the exact ground state energy of each subsystem, $-8.00$. The local basis is chosen by $\{ \ket{\psi_0^{(i)}}, A_\mu^{(i)} \ket{\psi_0^{(i)}} \}$ 
, where we take a site $\mu$ from the boundary of the $i$-th subsystem. With the local dimension $K=10$, the 2nd VQE requires $16$ qubits. Although the number of required qubits does not decreases here, note that the information of the whole Hilbert space is abandoned at the rate of $0.85$. We calculate the exact energy of $H^\mathrm{eff}$ instead of performing the 2nd VQE due to the computational cost. The resulting ground state energy is $-36.43$, which well reproduces the 
Lanczos
result $-36.76$ compared to the energy expectation value $H^\mathrm{eff}_{00}=-32.00$.

Let us discuss the concatenation up to the 3rd VQE for larger 2d systems which are difficult to classically simulate. We pick up 32-qubit ($L_x=8,L_y=4$) or 64-qubit ($L_x~L_y=8$) systems described by the Heisenberg Hamiltonian Eq. (\ref{eq_hamiltonian2d}). We show the subsystems at each step in Fig. \ref{fig3} (b) and (d). For the 32-qubit system, we consider two different choices of the local basis. After the 1st VQE on each $4$-qubit subsystem, the local excitation operators $\{ W_k^{(i)} \}_{k=1}^K$ are chosen from the identity and the Pauli operators at the boundary connected to other subsystems ($K=7$), leading to the 2nd VQE on the $6$ effective qubits. Next, we introduce two local operator sets:
\begin{align}\label{eq:local_basis_a}
    \text{(A): } \{ I, X_{\mu^\mathrm{eff}}^{(i)}, Y_{\mu^\mathrm{eff}}^{(i)}, Z_{\mu^\mathrm{eff}}^{(i)} \} 
\end{align}
where $\mu^\mathrm{eff}$ runs over the $6$ effective qubits ($K=19$), or
\begin{align}\label{eq:local_basis_b}
    \text{(B): } \{ I, X_{\mu,\mathrm{eff}}^{(i)},Y_{\mu,\mathrm{eff}}^{(i)}, Z_{\mu,\mathrm{eff}}^{(i)} \} 
\end{align}
where $\mu$ runs over the original boundary qubits  connected to other subsystems ($K=16$). Note that $A_{\nu,\mathrm{eff}}$ ($A=X,Y,Z$) physically represents the Pauli operator $A_\nu$ in the original system, but we should compute its matrix elements in the new basis for the effective model after the 1st VQE. Considering the result for the $32$-qubit system discussed later, we construct the local basis by Eq. (\ref{eq:local_basis_b}) after the 2nd VQE for the $64$-qubit system ($K=25$).

\begin{table}[t]
\begin{center}
\caption{
\label{Tab2}Numerical results for 2d Heisenberg anti-ferromagnetic systems. ``Deep VQE" indicates the results obtained by the proposed scheme. ``$\mathcal{W}_\mathrm{2nd}$" designates the set of local excitations from Eqs. (\ref{eq:local_basis_a}) and (\ref{eq:local_basis_b}) when we iterate up to the 3rd VQE. ``Local" is given by $H^{\rm eff}_{00}$ where $H^{\rm eff}$ is the effective Hamiltonian used for the last VQE in the protocol. ``Effective" means the Deep VQE results of the ground state energy under the assumption that VQEs are accurate enough, calculated by replacing the VQEs by the exact diagonalization.``Exact" means a ground state energy obtained by Lanczos method on  the quantum spin simulator QS$^3$\cite{2021arXiv210700872U} except for 64 sites, where -158.47* is obtained by the looper quantum Monte-Calro codes in ALPS\cite{ALBUQUERQUE20071187,Bauer_2011}. ``Qubits" indicates the number of qubits required through the protocol.}
\begin{tabular}{c|c|c||c|c|c|c}
\hline \hline
System  & Order & $\mathcal{W}_\mathrm{2nd}$ & Local & Effective & Exact & Qubits
\\
\hline
$16$ sites & 2nd & --- & $-32.00$ & $-36.43$ & $-36.76$ & 16
\\ \hline
$32$ sites & 2nd & --- & $-68.69$ & $-74.60$ & & 16
\\ \cline{2-5} \cline{7-7}
($8 \times 4$) & 3rd & (A) & $-68.55$  & $-69.57$ & $-76.30$ &20
\\ \cline{3-5} \cline{7-7}
 &    & (B) & $-68.55$ & $-71.49$ & & 16
\\ \hline
$64$ sites & 2nd & --- & $-147.03$ & $-153.11$ & $-158.47^*$  & 20
\\ \cline{2-5} \cline{7-7}
($8 \times 8$) & 3rd  & (B) & $-149.61$ & $ -151.39$ & & 16
\\
\hline \hline
\end{tabular}
\end{center}
\end{table}%

Table \ref{Tab2} shows the numerical results for the 2d systems. To evaluate the performance of the concatenated VQE with reducing the computational cost, we replace the VQEs by the exact diagonalization. We confirm that, for the $32$-qubit system, the 1st and the 2nd VQEs with hardware-efficient ansatz can reproduce the exact ground states of the corresponding subsystems with fidelity $1.000$ and $0.998$ respectively, indicating the validity of this replacement for assessing the Deep VQE results. We also simulate the concatenation up to the 2nd VQE in the way of Fig. \ref{fig3} (a) and (c) to compare the results. For the $32$-qubit system, while the 3rd VQE results are worse than that of the 2nd VQE due to repeated coarse-graining, they give approximate ground state energy $-69.57$ [for the local basis with (A)] and $-71.49$ [for the one with (B)], reproducing the Exact result $-76.30$ better than the local result $-68.55$. 

Let us discuss why the local basis choice (B) gives a better result than (A) to identify the better choice of the local basis when considering larger systems or further concatenation of VQEs. For the 3rd VQE, we employ the effective Hamiltonian after the 2nd VQE as a Hamiltonian of each subsystem. Since the effective Hamiltonian is generally nonlocal within each subsystem, it is difficult to describe excitations within each subsystem by a set of local operators. In our simulation for the 32-qubit system, the choice (A) captures local excitations in each subsystem while the choice (B) captures excitations that are local in the original system but nonlocal in each subsystem after the 2nd VQE. The better result of the choice (B) implies that the picture of linear excitations at the boundaries is maintained through the coarse-graining, and hence choosing the local excitation operators at the boundaries based on the original lattice is suitable also for further-concatenated VQEs or for larger systems. Based on this, we also simulate the $64$-qubit system with the local basis choice by (B). We obtain the 2nd VQE result $153.11$ and the 3rd VQE result $-151.39$, and both of them well reproduce the approximate value $-158.47$ computed by the looper quantum Monte-Calro codes in ALPS\cite{ALBUQUERQUE20071187,Bauer_2011}. While the concatenation up to the 3rd VQE gives a slightly worse upper bound for the  ground state energy than the 2nd VQE result, we can efficiently complete the simulation with further decreasing the size of quantum devices by $4$ qubits.

A better upper bound of the ground state energy or equivalently a more accurate value will be achieved if we consider local excitations near the boundaries of subsystems or higher order excitations when constructing the local basis also in two-dimensional systems, keeping the merit of decrease in qubits. Since low-entangled states in higher-dimensional systems are difficult to classically simulate by matrix-product-state-based methods such as the density matrix renormalization group \cite{DMRGreview}, the concatenation of VQEs will significantly benefit us in simulating classically-intractable higher dimensional systems.

\section{Conclusion and discussion}
We have proposed a DC method for the quantum-classical hybrid algorithm to 
solve a larger system with a small size of quantum computers. 
Specifically, VQE is performed recursively to reduce the physical dimensions,
while taking the interactions via the effective Hamiltonian.
Though we have only considered quasi-one-dimensional and two-dimensional Heisenberg anti-ferromagnetic models in the numerical simulations,
the proposed scheme is applicable to more complicated systems such as complex molecules such as molecular aggregates,  molecular crystals, and dendrimers.
If the subsystem is a strongly correlated system
which inevitably requires a highly entangled state only available by quantum computers,
the proposed scheme allows us to use quantum computers of relatively small size for large enough problems.

As a future direction, the proposed scheme can be hybridized with 
the classical tensor network approaches so that 
the effective Hamiltonian obtained from the first VQE can be solved by using tensor network methods as reported in Ref.~\cite{LeiWang19,Yuan20}.
More precisely, 
in Ref.~\cite{Yuan20},
the authors have proposed various types of quantum-classical hybrid tensor network models.
For example, if we employ an MPS ansatz classically after the first VQE in our proposal, the it can be seen as a connection of classical and quantum tensor networks mentioned in Ref.~\cite{Yuan20}.
Furthermore, the deep VQE, i.e., concatenation of VQEs at multiple stages is similar to quantum-quantum tensor network in Ref.~\cite{Yuan20}.
Specifically, if we employ the subspace search VQE to span a low energy subspace,
we have the same structure of tree type as that in Ref.~\cite{Yuan20}.
However, the local excitations and Schmidt process to span local bases cannot be regarded as a simple connection between two quantum tensor networks via classical tensor. Therefore it is useful for a quantum-classical hybrid tensor network to enhance its performance further by introducing a non-trivial classical processing on a classical part connecting different quantum tensors.

Note that, one of the reasons why the deep VQE works well for the ground state analysis of the $d(=1,2)$-dimensional antiferromagnetic Heisenberg model, despite introducing a dramatic reduction of degrees of freedom, is that the degrees of freedom $K$ for a local cluster automatically increases proportionally to the surface area of the cluster $O(\ell^{d-1})$ where $\ell$ is the length of one side of the cluster.
We should emphasize that such area law of space expansion employed in the deep VQE partially incorporates the property of entropic area law~\cite{Calabrese_2004} for  a $d$-dimensional quantum many-body system consisting only of short-range interactions that the entanglement entropy is proportional to the surface area $O(l^{d-1})$ of the subsystem.
One guiding principle for deepening the deep VQE based of the entropic area law is to develop a procedure that controls the number of local degrees of freedom of the cluster to be increased by an arbitrary order of integer power with respect to its surface area, so that as a extreme case of the procedure the local degrees of freedom can be increased by $O(\exp(l^{d-1}))$ satisfying the entropic area law.
For example, taking into account an effect of a $n(>1)$th-order perturbation, it would be a naive extension of the deep VQE to prepare a local degree of freedom proportional to the $n$th power of the surface area by letting the $n$ bodies Pauli products act on the qubits near the interface with respect to the ground state of the cluster. The validity of such an extension originating from this work is one of the future issues.

\section*{Acknowledgement}
KF is supported by JST ERATO JPMJER1601, and JST CREST JPMJCR1673. This work is supported by MEXT Quantum Leap Flagship Program (MEXT Q-LEAP) Grant Number JPMXS0118067394 and JPMXS0120319794. Kaoru Mizuta appreciates the support of WISE Program from MEXT and a Reseach Fellowship for Young Scientists from JSPS (No.20J12930). WM wishes to thank JSPS KAKENHI No.\ 18K14181 and JST PRESTO No.\ JPMJPR191A.
HU is supported by KAKENHI Nos. 17K1435, 21H04446 and 21H05191, and JST PRESTO No. JPMJPR1911, and the COE research grant in computational science from Hyogo Prefecture and Kobe City through Foundation for Computational Science.
We are grateful for allocating computational resources of the HOKUSAI BigWaterfall supercomputing system at RIKEN.

\bibliographystyle{apsrev4-1}
\input{output.bbl}

\end{document}

%% file: output.bbl
%